\documentclass[aps,prd,showpacs,%
preprint,
preprintnumbers,superscriptaddress,nofootinbib,english]{revtex4}
\usepackage{graphicx}
\providecommand{\tabularnewline}{\\}
\providecommand{\crtbt}{Centre de Recherche sur les Tr\`es Basses Temp\'eratures, %
SPM-CNRS, BP 166, 38042 Grenoble, France}
\providecommand{\csnsm}{Centre de Spectroscopie Nucl\'eaire et de Spectroscopie de Masse,%
IN2P3-CNRS, Universit\'e Paris XI, bat 108, 91405 Orsay,  France}
\providecommand{\iek}{Institut f\"ur Experimentelle Kernphysik, Universit\"at Karlsruhe (TH), %
Gaedestr. 1, 76128 Karlsruhe, Germany}
\providecommand{\fzk}{Forschungszentrum Karlsruhe, Institut f\"ur Kernphysik, %
Postfach 3640, 76021 Karlsruhe, Germany}
\providecommand{\drecam}{CEA, Centre d'\'Etudes Nucl\'eaires de Saclay,
DSM/DRECAM, 91191 Gif-sur-Yvette Cedex, France}
\providecommand{\dapnia}{CEA, Centre d'\'Etudes Nucl\'eaires de Saclay,
DSM/DAPNIA, 91191 Gif-sur-Yvette Cedex, France}
\providecommand{\ipnl}{Institut de Physique Nucl\'eaire de Lyon - Universite Claude Bernard Lyon~1, IN2P3-CNRS, 4 rue Enrico Fermi, 69622 Villeurbanne Cedex, France}
\providecommand{\iap}{Institut d'Astrophysique de Paris, UMR7095 CNRS,
Universit\'e Pierre et Marie Curie, 98 bis Bd Arago, 75014 Paris, France}

\usepackage{babel}

\begin{document}

\preprint{LYCEN 2006-10}

\title{Measurement of the response of heat-and-ionization germanium
detectors to nuclear recoils}

\author{A. Benoit}
\affiliation{\crtbt}
\author{L. Berg\'e}
\affiliation{\csnsm}
\author{J.~Bl$\mbox{\"u}$mer}
\affiliation{\iek}
\affiliation{\fzk}
\author{A. Broniatowski}
\affiliation{\csnsm}
\author{B. Censier}
\affiliation{\csnsm}
\author{A. Chantelauze}
\affiliation{\fzk}
\author{M. Chapellier}
\affiliation{\drecam}
\author{G. Chardin}
\affiliation{\dapnia}
\author{S. Collin}
\affiliation{\csnsm}
\author{X. Defay}
\affiliation{\csnsm}
\author{M. De J\'esus}
\affiliation{\ipnl}
\author{H.~Deschamps}
\affiliation{\dapnia}
\author{P. Di Stefano}
\affiliation{\ipnl}
\author{Y. Dolgorouky}
\affiliation{\csnsm}
\author{L. Dumoulin}
\affiliation{\csnsm}
\author{K. Eitel}
\affiliation{\fzk}
\author{M. Fesquet}
\affiliation{\dapnia}
\author{S. Fiorucci}
\affiliation{\dapnia}
\author{J. Gascon}
\affiliation{\ipnl}
\author{G. Gerbier}
\affiliation{\dapnia}
\author{C. Goldbach}
\affiliation{\iap}
\author{M. Gros}
\affiliation{\dapnia}
\author{M. Horn}
\affiliation{\fzk}
\author{A. Juillard}
\affiliation{\csnsm}
\author{R. Lemrani}
\affiliation{\dapnia}
\author{A. de Lesquen}
\affiliation{\dapnia}
\author{M. Luca}
\affiliation{\ipnl}
\author{S. Marnieros}
\affiliation{\csnsm}
\author{L. Mosca}
\affiliation{\dapnia}
\author{X.-F.~Navick}
\affiliation{\dapnia}
\author{G. Nollez}
\affiliation{\iap}
\author{E. Olivieri}
\affiliation{\csnsm}
\author{P. Pari}
\affiliation{\drecam}
\author{V. Sanglard}
\affiliation{\ipnl}
\author{L. Schoeffel}
\affiliation{\dapnia}
\author{F. Schwamm}
\affiliation{\dapnia}
\author{M. Stern}
\affiliation{\ipnl}
\collaboration{The EDELWEISS Collaboration}
\noaffiliation

\date{\today}

\begin{abstract}
The heat quenching factor $Q'$ (the ratio of the heat signals produced
by nuclear and electron recoils of equal energy) of the heat-and-ionization
germanium bolometers used by the EDELWEISS collaboration 
has been measured.
It is explained how this factor affects the energy scale and the
effective quenching factor observed in calibrations with
neutron sources.
This effective quenching effect is found to be equal to $Q/Q'$, where $Q$
is the quenching factor of the ionization yield. 
To measure $Q'$, a precise EDELWEISS measurement of $Q/Q'$
is combined with values of $Q$ obtained from a review of all available 
measurements of this quantity in tagged neutron beam experiments.
The systematic uncertainties associated with this method to evaluate $Q'$
are discussed in detail.
For recoil energies between 20 and 100 keV, the resulting heat 
quenching factor is $Q'$ = 0.91~$\pm$~0.03~$\pm$~0.04, 
where the two errors are the contributions from
the $Q$ and $Q/Q'$ measurements, respectively.
The present compilation of $Q$ values and evaluation of $Q'$Ê
represent one of the most precise determinations of the absolute energy
scale for any detector used in direct 
searches for dark matter.

\end{abstract}

\pacs{29.40.Wk, 95.35.+d}
%29.40.Wk, %Solid-state detectors
%95.35.+d, % Dark Matter (stellar, interstellar, galactic and cosmological)

\maketitle

\section{Introduction}

Heat-and-ionization germanium detectors
are extensively used for direct search of the weakly interacting massive
particles (WIMPs) that could constitute the Dark Matter halo of our
Galaxy\cite{cdms-soudan,edw2003}. 
The scattering of a WIMP on a Ge atom produces a nuclear recoil
with a kinetic energy in the ten's of keV range\cite{bib-lewin}.
The recoil is stopped in the detector volume within a distance of 100 nm. 
The energy loss occurs as a combination of ionization (electronic
dE/dx) and atomic collisions (nuclear dE/dx)\cite{bib-lindh}. 
In heat-and-ionization detectors, the event is identified using two 
signatures.
The first is the ionization signal, corresponding to the collection on 
electrodes of the electron-hole pairs created by the energy loss process. 
The second is the heat (or phonon) signal, recorded by a thermal sensor 
in contact with the germanium crystal. 
It has been demonstrated that combining the two simultaneous 
measurements provides an efficient discrimination against the large 
background of electron recoils originating from the natural $\gamma$ 
and $\beta$ radioactivity\cite{cdms-soudan,edw2003}. 
The basis of this discrimination is that the number of electron-hole 
pairs created by an electron recoil of a given energy is three to four 
times larger than that created by a nuclear recoil of the same energy.

The process at the origin of the reduced ionization yield has
been extensively studied since the original work of Lindhard\cite{bib-lindh}.
The effect has been measured repeatedly as a function of recoil
energy in germanium ionization detectors
at liquid Nitrogen temperature (77 K)%
~\cite{bib-chasman1,bib-sattler,bib-chasman2,bib-jones,bib-messous,bib-baudis}, 
and more recently at 35~mK~\cite{bib-sicane}.
In these experiments, the ionization signals recorded with such detectors
are first calibrated using gamma-ray sources, producing energetic 
electron recoils. 
This energy scale is called keV-equivalent-electron (keV$_{ee}$).
Using this calibration, the detector is then exposed to a neutron source.
The elastic collisions of neutrons with atoms in the detector volume
produce nuclear recoils. 
The energy of the recoils is constrained by the use 
of monoenergetic neutron beams, the detection 
of the neutron scattering angle and/or its time-of-flight%
~\cite{bib-chasman1,bib-sattler,bib-chasman2,bib-jones,bib-messous,bib-baudis}.
It is then observed that the ionization yield
for a nuclear recoil is a factor $Q\sim0.25$ smaller than that produced
by an electron recoil of equal energy. 

In analogy with the ionization quenching factor $Q$, one can similarly 
introduce a quenching factor $Q'$ for the heat signal in thermal detectors. 
This factor does not affect the ability
of heat-and-ionization detectors to discriminate electron and
nuclear recoils\footnote{Except in the very special case
where $Q'$=$Q$, clearly excluded by the measurements discussed
in the following.}. 
However, this factor enters in the determination of the 
energy calibration for nuclear recoils and, consequently,
in the energy threshold for their detection.
Heat-and-ionization detectors are designed so that the initial
deposited energy thermalizes in time for the signal read-out, 
independently of the process at its origin, with
as little as possible losses to the outside world. 
It is thus expected that $Q'$ is close to one. 
Possible deviations can only arise from processes that affect 
differently electron and nuclear recoils.
In Ge crystal at cryogenic temperatures (typically 10 to 100 mK),
the possible sources of such differences are usually considered to be small,
but have never been measured precisely. Examples of such processes
are the storage of energy in stable crystal defects generated by the
recoiling nuclei, or the emission of photons during the initial stage
of ionization. Systematic detector-dependent effects could also appear
if the heat signal is read out before allowing for the full thermalization
of the phonon excitations in the crystal.

\begin{figure}
\begin{center}\includegraphics[scale=0.5]{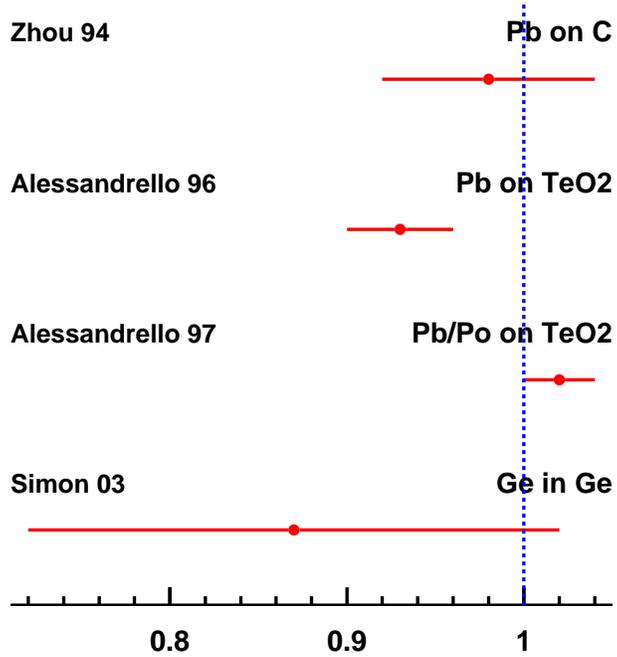}\end{center}
\caption{\label{cap:previous}Available measurements of $Q'$, the quenching
of the heat signal for nuclear recoils, in germanium, diamond and
TeO$_{2}$. The measurements are from Refs. \cite{bib-zhou} (Zhou
94), \cite{bib-aless} (Alessandrello 96 and 97) and \cite{bib-sicane}
(Simon 03). In all cases, the recoil kinetic energies are approximately
100 keV.}
\end{figure}

A direct measurement of $Q'$ is a delicate experiment. Fig.~\ref{cap:previous}
summarizes the available measurements for all type of low-temperature devices. 
They are all consistent with unity, although the only measurement directly relevant to
germanium, performed  with a tagged neutron beam, is not very precise~\cite{bib-sicane}.
The measurements in the other bolometric devices involve $^{206}$Pb 
recoils from $^{210}$Pb alpha decays close to the surface
of diamond~\cite{bib-zhou} and TeO$_{2}$~\cite{bib-aless} detectors. 
It could be argued that processes that could lead to deviations from
unity in a germanium semiconductor might not show up as strongly,
or even cancel out, in the other measured substrates. The technique
of using surface recoils cannot be applied easily to germanium heat-and-ionization
detectors because of the presence of electrodes, dead layers and
systematic effects in the drift of electrons and
holes near the surface of the detector\cite{bib-surf}.

However, as it will be shown here, heat-and-ionization bolometers
provide an accurate measurement of the ratio $Q/Q'$. 
This idea has already been qualitatively stated in Ref.~\cite{bib-shutt}, 
where it was observed that the data appeared to be consistent with
$Q'=1$, without defining an explicit procedure for extracting a
quantitative value.
In this work, we define a procedure in which the $Q'$
value for Ge recoils in Ge is obtained by combining the measurements
of $Q/Q'$ from heat-and-ionization detectors with the available direct
$Q$ measurements. In the first section, the mathematical formulae
necessary to obtain the final result and, more importantly, its systematic
uncertainty, are presented. In the second section, we review the available
$Q$ data, compute their average values as a function of recoil energy,
and evaluate their statistical and systematic uncertainties. The third
section presents the results of $Q/Q'$ measurements using the EDELWEISS
detectors, and discusses their associated systematic errors. The results
are also compared to those of Ref.~\cite{bib-shutt}. In the fourth
section, the $Q/Q'$ measurements are compared with the computed average
$Q$ values. The final result can be interpreted either as the most
precise available measurement of the heat quenching factor $Q'$ for
Ge atoms recoiling in the bulk of a Ge crystal, or, conversely, as
an evaluation of the uncertainties on the absolute energy scale of nuclear
recoils in the EDELWEISS heat-and-ionization detectors due to imperfection
of the understanding of the underlying physics.

\section{Definition of terms}

\subsection{Ionization measurement\label{sub:Ion-meas}}

The amplitude of the ionization signal $A_{I}$ is proportional to
the number of collected electron-hole pairs $N$. In heat-and-ionization
detectors, the applied difference of potential on the electrodes $V$
is chosen large enough so that losses due to trapping are small. This
is verified%
\footnote{In EDELWEISS detectors, this is verified for $|V|$ $>$ $\sim$3
Volts. %
} by observing that $A_{I}$ is independent of $V$. Under these conditions,
for electron recoils\[
A_{I,\gamma}\propto N_{\gamma}=\frac{E}{\epsilon_{\gamma}}\]

where $E$ is the electron recoil energy and $\epsilon_{\gamma}$
is the average energy necessary to create one electron-hole pair.
At 77 K, $\epsilon_{\gamma}$= 2.96 eV\cite{bib-knoll}. At cryogenic
temperature, this value is expected to increase slightly due to the
variation of the gap energy. A value of 3.0 $\pm$0.1 eV is adopted
here, consistent with the measurement of Ref.~\cite{bib-epsgamma} and
the value used by the EDELWEISS and CDMS experiments.

The amplitude $A_{I}$ is calibrated using gamma-ray sources to provide
$E_{I}$, the energy in units of keV$_{ee}$.

For nuclear recoils, the average energy per electron-hole pair $\epsilon_{n}$
is approximately 12 eV, and varies with energy. The quenching factor
for the ionization signal of nuclear recoils is defined as $Q$ =
$\epsilon_{\gamma}/\epsilon_{n}$, and\[
A_{I,n}\propto N_{n}=\frac{E}{\epsilon_{n}}=Q\frac{E}{\epsilon_{\gamma}}\]

The Lindhard theory describing the energy losses of energetic ions
in matter provides some predictions for the value of $Q$\cite{bib-lindh}.
Within this model, the fraction of the incident energy dissipated
in ionization is calculated from the electronic and nuclear stopping
power $dE/dx$ of Ge ions in Ge. Because of the approximations behind
this calculation, it cannot be expected to provide an accurate prediction
of the absolute value of $Q$. However, it will be shown in the following
section that it describes reasonably well the overall energy dependence
of $Q$. The Lindhard model will be used to interpolate between measurements
of $Q$ at different energies. We thus introduce this prediction as
a reference function, $Q_{ref}(E)$, evaluated for Ge recoils in Ge
following the usual parametrizations~\cite{bib-lewin,bib-lindh}:
\begin{eqnarray}
Q_{ref} & = & \frac{kg(\epsilon)}{1+kg(\epsilon)}\label{eq:lind}\end{eqnarray}
 where $\epsilon$ is a dimensionless energy, $k$ is related to the
electronic energy loss ($dE/dx_{electron}=k\sqrt{\epsilon}$), and
for Ge recoils in Ge: \begin{eqnarray}
\epsilon & = & \frac{{11.5}}{Z^{7/3}}E\;=\;0.00354\; E(\mbox{keV})\label{eq:lind1}\\
k & = & 0.133\frac{{Z^{2/3}}}{A^{1/2}}\;=\;0.157\label{eq:lind2}\\
g(\epsilon) & = & 3\epsilon^{0.15}+0.7\epsilon^{0.6}+\epsilon\label{eq:lind3}\end{eqnarray}

\subsection{The heat or phonon measurement}

For electron recoils in heat-and-ionization detectors, the heat
signal amplitude $A_{H}$ is due to the initial energy of the recoil $E$ and
to the Joule heating associated to the current of collected electron
and holes through the detector (Neganov-Luke effect\cite{bib-luke}):\begin{equation}
A_{H,\gamma}\;\propto\; E+N_{\gamma}eV\;=\;(1+eV/\epsilon_{\gamma})E\label{eq:ah}\end{equation}

where $e$ is the charge of the proton and $V$ is the absolute value
of the applied potential. For simplicity, we note the quantity $eV/\epsilon_{\gamma}$
as $v$, and $A_{H,\gamma}\propto(1+v)E$. For nuclear recoils in a truly
calorimetric detector ($Q'$=1), the amplitude is:\[
A_{H,n}(Q'=1)\;\propto\; E+N_{n}eV\;=\;(1+vQ)E\]

If the phonon yield for nuclear recoils is reduced by a factor $Q'$,
the Joule heating contribution remains proportional to the number
of created electron-hole pairs $N$ and:\[
A_{H,n}\;\propto\;(Q'+vQ)E\]

Once the heat signal amplitude $A_{H,\gamma}$ 
is calibrated in units of keV$_{ee}$ using a
gamma-ray source, we obtain the energy $E_{H,\gamma}=E$ and\[
E_{H,n}\;=\;\frac{Q'+vQ}{1+v}E\]

\subsection{Recoil energy measurement for $Q'$=1 and $Q'$$\neq$1}

\label{sub:ydef}The two independent quantities $E_{H}$ and $E_{I}$
can be used to evaluate the following two quantities:\begin{equation}
x=(1+v)E_{H}-vE_{I}\label{eq:defx}\end{equation}
\begin{equation}
y=E_{I}/x\label{eq:defy}\end{equation}

We also define $(x_{\gamma},y_{\gamma})$ and $(x_{n},y_{n})$ as
the values of $(x,y)$ in the case of electron and nuclear recoils,
respectively. In the case where $Q'=1$, the variable $x$ and $y$
have simple interpretations. As $x_{\gamma}=x_{n}=E$, the variable
$x$ represents the recoil energy, irrespective of the type of recoil.
The variable $y$ measures the quenching effect relevant to the incident
particle, as $y_{\gamma}=1$ for electron recoils, while $y_{n}=Q$
for nuclear recoils. 

In the more general case where $Q'\neq1$, the relationships for electron
recoils $x_{\gamma}=E$ and $y_{\gamma}=1$ are still true. For nuclear
recoils, however, the following values are obtained:\[
x_{n}=Q'E\]
\[
y_{n}=Q/Q'\]

A plot of the measured values of $y$ versus $x$, as it is typically
done\cite{cdms-soudan,edw2003}, will still display the characteristic
separation of the two populations of electron and nuclear recoils.
However the position of the so-called {}``nuclear recoil band''
($y(x)=y_{n}(x)$) has to be interpreted with care. First, the measured
$y_{n}$ values will be divided by a factor $Q'$ relative to
the true value of $Q$. This has no consequence for the identification 
of nuclear recoils in Refs.\cite{cdms-soudan,edw2003}, since the position
of this band is always taken from the results of the neutron source
calibrations of the relevant detector. However, the apparent recoil
energy for nuclear recoil $x_{n}$, obtained by assuming $Q'=1$,
is not equal to the true recoil energy $E$. For example, $Q'$=0.9,
and an apparent recoil energy threshold of $x_{n}$= 10 keV corresponds
to a true threshold of $E$ = 11 keV.

The neutron calibration of the first heat-and-ionization detectors\cite{bib-shutt}
quickly established that $Q'$ was not very different from 1, as the
measured $y_{n}$ values were compatible to the direct measurement
of $Q$ performed at 77 K with Ge diode detectors. This work pursues
this idea further by actually performing the evaluation of\begin{equation}
Q'=\frac{Q}{y_{n}}\label{eq:qsury}\end{equation}

using a compilation of the available direct measurements for $Q$
and the most precise $y_{n}$ measurements performed by the EDELWEISS
collaboration with its detectors. A complete assessment of the statistical
and systematic uncertainties associated to both types of measurement
will also be performed.

\subsection{Propagation of uncertainties}

To evaluate the systematic uncertainties on $Q'$ as evaluated with
Eq.~\ref{eq:qsury}, it is useful to study its dependence on different
variables. First, the uncertainty on $Q'$ depends linearly on those
on the $Q$ measurement. In the next section, these uncertainties
are discussed in details. As for the uncertainties on $y_{n}$, we
have:

\begin{eqnarray}
\frac{\partial y_{n}}{\partial E_{H}} & = & -\frac{(1+v)Q}{Q'^{2}E}\label{eq:dydeh}\\
\frac{\partial y_{n}}{\partial E_{I}} & = & \frac{(Q'+vQ)}{Q'^{2}E}\label{eq:dydei}\\
\frac{\partial y_{n}}{\partial v} & = & -\frac{Q\,(Q'-Q)}{Q'^{2}(1+v)}\label{eq:dydv}\end{eqnarray}

An effect that must be taken into account when evaluating the ratio
of the $Q(E)$ and $y_{n}(x)$ measurements is that in general $x\neq E$.
At lowest order (small deviations of $Q'$ from 1 and small slope
$dQ/dE$), the value of $y_{n}(E)$ can be obtained from that of $y_{n}(x)$
by:\begin{equation}
y_{n}(E)\,=\, y_{n}(x)+\frac{\partial Q}{\partial E}\frac{(1-Q')}{Q'^{2}}E\label{eq:ecor}\end{equation}

where, as it will be shown later, $dQ/dE$ can be taken from the Lindhard
theory with a good precision.

\section{A review of Q measurements}

\label{sec:Qrev}
\begin{figure}
\begin{center}\includegraphics[scale=0.8]{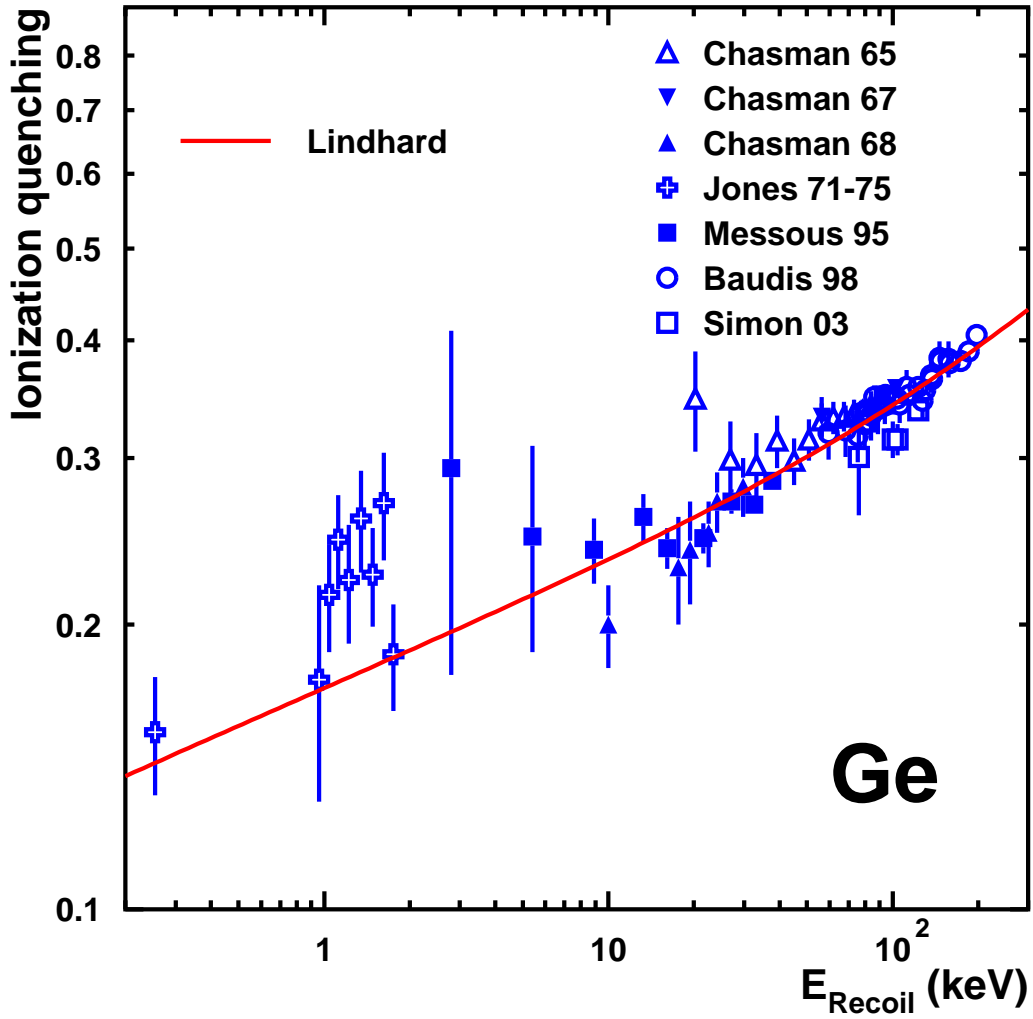}\end{center}
\caption{\label{cap:qrev}Experimental results of the direct measurement of
the ionization quenching for germanium recoils in germanium, from
Refs. \cite{bib-chasman1} (Chasman 65), \cite{bib-chasman2} (Chasman
67 and 68), \cite{bib-jones} (Jones 71 and 75), \cite{bib-messous}
(Messous 95), \cite{bib-baudis} (Baudis 98) and \cite{bib-sicane}
(Simon 03). The line represents Eq.~\ref{eq:lind}, with parameter
values as of Eqs.~\ref{eq:lind1} to~\ref{eq:lind3}.}
\end{figure}

Fig.~\ref{cap:qrev} shows the results%
\footnote{%
The data from Ref.~\cite{bib-sattler} are not included in this review
because of the reasons explained in Ref.~\cite{bib-chasman2}. %
} of the direct measurements of ionization quenching for germanium recoils in germanium
from Refs. \cite{bib-chasman1,bib-chasman2,bib-messous,bib-baudis,bib-jones,bib-sicane}.
The measurements from Ref.~\cite{bib-shutt} are discussed in Sect.~\ref{sec:Q/Q'},
as they consist in a measurement of $Q/Q'$ and not of $Q$. No large
inconsistency can be observed. The recoil energy dependence of the
data is well described by the Lindhard model (Eq.~\ref{eq:lind}),
shown as a continuous curve on the same plot.

The measurements were performed with various experimental techniques.
The systematic uncertainty on $Q$ can be reduced by combining them.
However, the recoil energy intervals used in the measurements are
not the same, and one must take into account the energy dependence
of $Q$ when combining two measurements performed at different average
recoil energies. For this reason, we will not perform weighted averages
of $Q(E_{R})$, but instead use $Q(E_{R})/Q_{ref}(E_{R})$, with $Q_{ref}(E_{R})$
given by Eq.~\ref{eq:lind}. These weighted averages will be done
for each of the recoil energy bin used in Sect.~\ref{sec:Q/Q'}.
They are shown in Table~\ref{tab:q:}. In order to take into account
possible correlations between the error bars within the same experimental
data set in a given energy interval, only one measurement per experiment
is used per interval. This measurement is chosen as the one with the
smallest error. The weights are the square of the inverse of the quoted
total uncertainties. The uncertainty on the combined $Q(E_{R})/Q_{ref}(E_{R})$
ratios is computed assuming that the errors are not correlated between
the measurements. If in a given energy bin, the reduced chi-square
of the combination of $N$ results, $\chi^{2}/(N-1)$, is larger than
one, the uncertainty is multiplied by the square root of this quantity.
The average recoil energy $<E_{R}>$ for each energy bin is calculated
as the energy of the individual measurements averaged using the same
weights. Finally, the average ratio $Q(E_{R})/Q_{ref}(E_{R})$ is
multiplied with $Q_{ref}(<E_{R}>)$, yielding a set of values of $Q(<E_{R}>)$
with their experimental uncertainties that exploits at best the available
data sets.

Table~\ref{tab:q:} shows that the different measurements are consistent
within small adjustment to the quoted errors, with four out of eight
reduced $\chi^{2}$values above unity, of which two are slightly above
2. This consistency is remarkable considering that the measurements
were performed over a period of 38 years, with experimental setups
and techniques that differ considerably and with constant improvements
in the understanding of the systematic biases. Over the range from
5 to 200 keV, there is no sign of deviation of $Q/Q_{ref}$ larger
than $\sim$$\pm$5\%. This indicates that the choice of the function
$Q_{ref}(E_{R})$ for interpolating the results within an energy bin
is appropriate.

In the energy bin from 80 to 100 keV, the results of the three available
measurements are consistent within less than 1\%. The combined result
has an uncertainty of 1\%, making it one of the most precisely measured
quenching factor in any type of detector. One can also note that the
absolute value of $Q_{ref}(E_{R})$ is very close to the experimental
measurements. This is not too surprising since the overall normalizations
of the nuclear and electronic stopping powers in the Lindhard model
(Eqs.~\ref{eq:lind1} to~\ref{eq:lind3}) are based on experimental
data. However the experimental values of $Q(<E_{R}>)$ do not depend
on this precise choice of reference.

\section{The EDELWEISS Q/Q' measurement}

\label{sec:Q/Q'}As discussed in Sect.~\ref{sub:ydef}, the EDELWEISS
heat-and-ionization detectors provide a measurement of $y(x)$, where
$x=Q'E$ and, for events due to the elastic scattering of neutrons
on Ge nuclei, the $y(x)$ measurements have a Gaussian dispersion
centered on the value $y_{n}$= $Q/Q'$. In this section, we present
the results obtained with five detectors%
\footnote{The two other EDELWEISS detectors\cite{edw-calib}
GeAl6 and GeAl9 have not been used because of the stringent requirements
for resolution and precision of the energy calibration necessary for the 
present measurement.%
}, labeled GeAl10, GGA1, GGA3,
GSA1 and GSA3. Their characteristics and performances are detailed
in Ref.~\cite{edw2003,bib-edw2002,edw-calib}, where the experimental
setup and its operation are also described. The main difference between
these detectors is whether the amorphous layer under the aluminum
electrodes is made of germanium (GGA series) or of silicon (GSA series),
or whether there is none (GeAl series). This difference affects the
charge collection properties of the detector~\cite{edw2003,bib-edw2002}.
The electrodes are polarized to a potential of 4.00 V, resulting in
a value of $v$ = 4/3.

The experimental procedure and data analysis proceed according to
the following steps. First, the detectors are calibrated in keV$_{ee}$
using $\gamma$-ray sources (Sect.~\ref{sub:gcal}). They are then
exposed to a fast neutron flux of the order of 1 to 10 n/cm$^{2}$/h
using a weak $^{252}$Cf source. From this data is deduced $y_{n}$,
the average value of $y$ as a function of $x$ for elastic neutron
scattering events (Sect.~\ref{sub:ncal}). At this point, the results
are compared with those of Ref.~\cite{bib-shutt}. The data are then
corrected for multiple scattering and energy shifts (Sect.~\ref{sub:corrections})
and the systematic uncertainties on the measurement are evaluated
(Sect.~\ref{sub:Syst}). Finally, the quenching of the heat signal
$Q'$ is obtained by dividing the $Q$ values of Sect.~\ref{sec:Qrev}
by the corrected $y_{n}$ values.

\subsection{Calibration with $\gamma$ rays}

\label{sub:gcal}The first step is the calibration of the detectors
in keV$_{ee}$ with $\gamma$-ray sources. An often overlooked difficulty
in quenching factor measurements is that of the precise calibration
of the detector in the low-energy range relevant for dark matter searches
(10 to 30 keV, and even below for quenched signals). Eqs.~\ref{eq:dydeh}
and~\ref{eq:dydei} show that, for $Q'\sim1$ and $v=4/3$, a 1\%
precision on $y_{n}$ requires a precision of 1.3\% and 0.7\% on the
measurement of the heat and ionization signals, respectively. 

\begin{figure}
\begin{center}\includegraphics[scale=0.8]{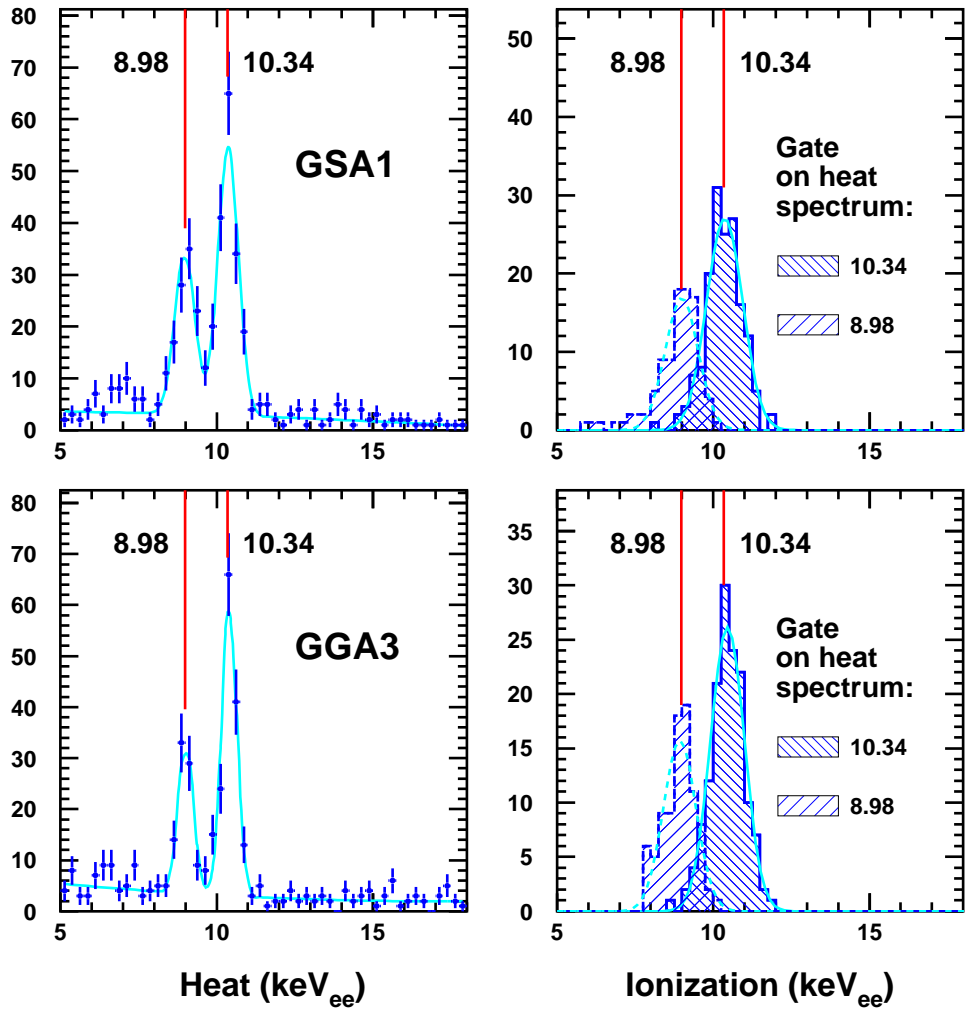}\end{center}
\caption{\label{cap:tstcal}Example experimental checks of the calibration
at 10 keV$_{ee}$ of the EDELWEISS detectors GSA1 (top) and GGA3 (bottom). 
The full lines indicate the expected
position of the activation peaks at 8.98 and 10.34 keV. 
Left: heat signal spectra. The FWHM resolution of the peaks are
 0.76$\pm$0.05 and 0.58$\pm$0.05 keV for GSA1 and GGA3, respectively. 
Right: Ionization spectra. To improve the identification of the two peaks,
these spectra were obtained using gates on the 8.98 and 10.34 peaks in
the heat spectra on the left.
The FWHM resolution of the ionization peaks are 1.3$\pm$0.1 keV for 
both detectors.}
\end{figure}

It is difficult to calibrate the response of the bulk of a detector
to $\sim$10 keV $\gamma$ rays because of their short attenuation
length, which excludes the possibility of using external sources.
This difficulty is somehow alleviated in semi-conductor detectors.
Because of the good linearity of their ionization signal, the extrapolation
of the calibration from higher energy is more reliable than, for example,
scintillating detectors. In the case of germanium detectors in an
underground and very-low background environment, there is the additional
unique possibility to check the calibration at 10 keV$_{ee}$ using
the doublet arising from the slow decay of the cosmogenic isotopes
$^{68}$Ge and $^{65}$Zn ($T_{1/2}$ = 271 and 244 days, respectively),
and also the isotope $^{71}$Ge arising from the activation of the
detector with the $^{252}$Cf neutron source ($T_{1/2}$ = 11.5 days).
Fig.~\ref{cap:tstcal} shows the two peaks at 8.98 and 10.34 keV
coming from the electron conversion decay of Ge and Zn, respectively,
as recorded in EDELWEISS detectors in a low-background run following
a neutron calibration. With these data, it is possible to control
with a precision of 1\% the absolute energy calibration at 10 keV$_{ee}$
of the ionization and heat signals independently. At other energies,
the relative calibration of the two signals can be easily checked
using the distribution of the ratio of the two signals recorded in
the Compton plateau produced by a $^{137}$Cs $\gamma$-ray source.

\subsection{Neutron calibration data}

\label{sub:ncal}
\begin{figure}
\begin{center}\includegraphics[scale=0.8]{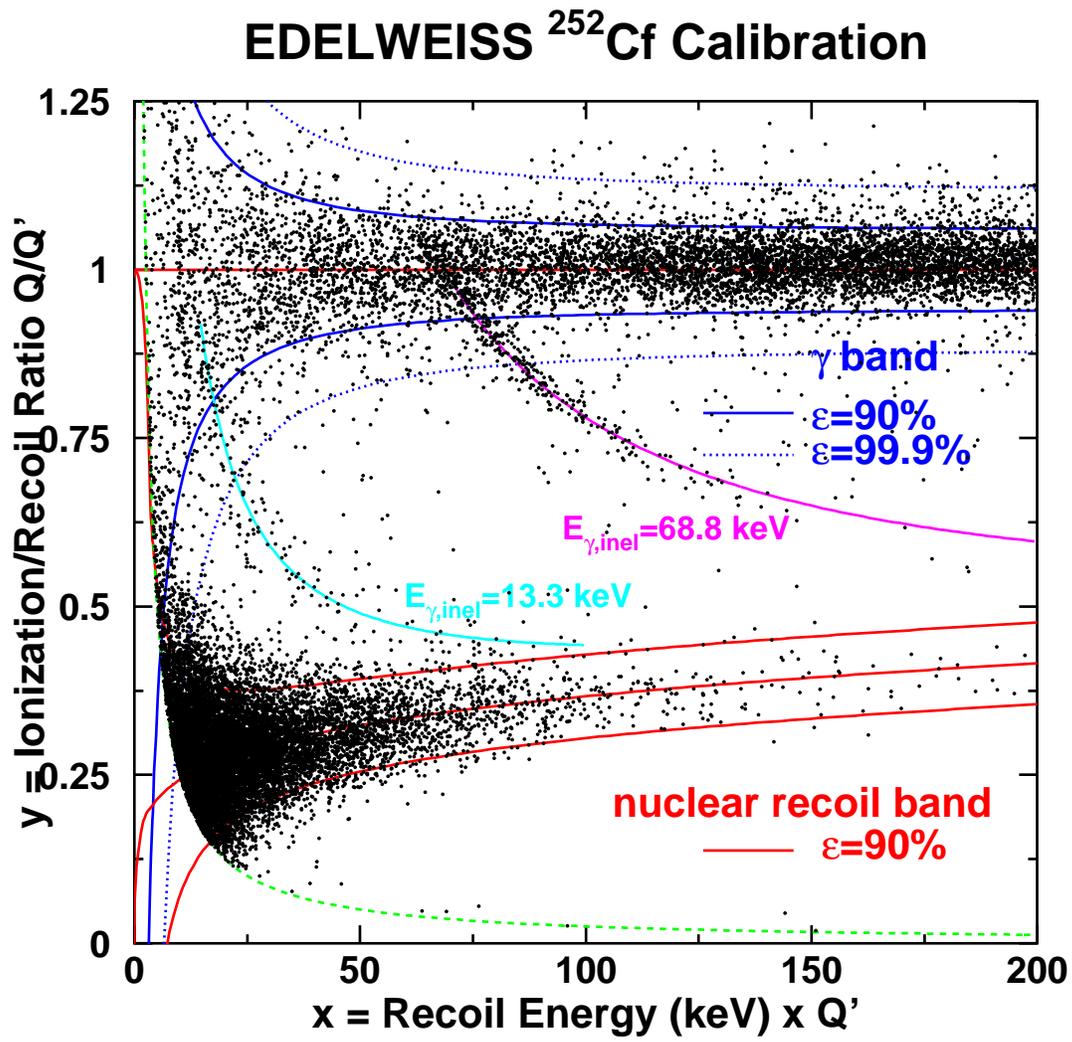}\end{center}
\caption{\label{cap:yvsx}Experimental data recorded with a $^{252}$Cf neutron
source with the EDELWEISS detectors GSA3, GSA1 and GGA3. }
\end{figure}

Fig.~\ref{cap:yvsx} shows the distribution of $y(x)$
recorded in the detectors GSA1, GSA3 and GGA3 when exposed to the
$^{252}$Cf neutron source. The data of these three detectors are
combined in that plot because they have similar experimental resolutions.

Of the 20340 counts in Fig.~\ref{cap:yvsx}, the overwhelming majority
is part of four well-understood populations, described here. The first
two populations are by far the most important. The first, centered
at $y=y_{\gamma}=1$, corresponds to $\gamma$ rays. It is well contained
within the 90\% efficiency zone computed from the experimental resolution
on the heat and ionization channels for these detectors, also shown
on that figure. The second population lies mostly inside the zone
expected for Ge nuclear recoils. On Fig.~\ref{cap:yvsx}, the center
of this zone is parametrized as in Ref.~\cite{edw2003} as a band
centered at $y_{ref}$ = $\alpha x^{\beta}$, with $\alpha=0.16$
and $\beta=0.18$. Two lesser populations are observed, corresponding
to the inelastic collision of a neutron on a $^{73}$Ge nucleus with
the excitation of the 13.3 and 68.8 keV states. In these cases, the
value of $y$ is the energy-weighted average of the quenching of the
Ge recoil and of the electromagnetic radiation emitted in coincidence.
Inelastic excitation of states decaying with higher-energy $\gamma$
rays, observed as Compton electrons in the detector, would produce
counts evenly spread between the two zones associated with electron
and nuclear recoils, and over the entire energy range displayed in
Fig.~\ref{cap:yvsx}. 

\begin{figure}
\begin{center}\includegraphics[scale=0.8]{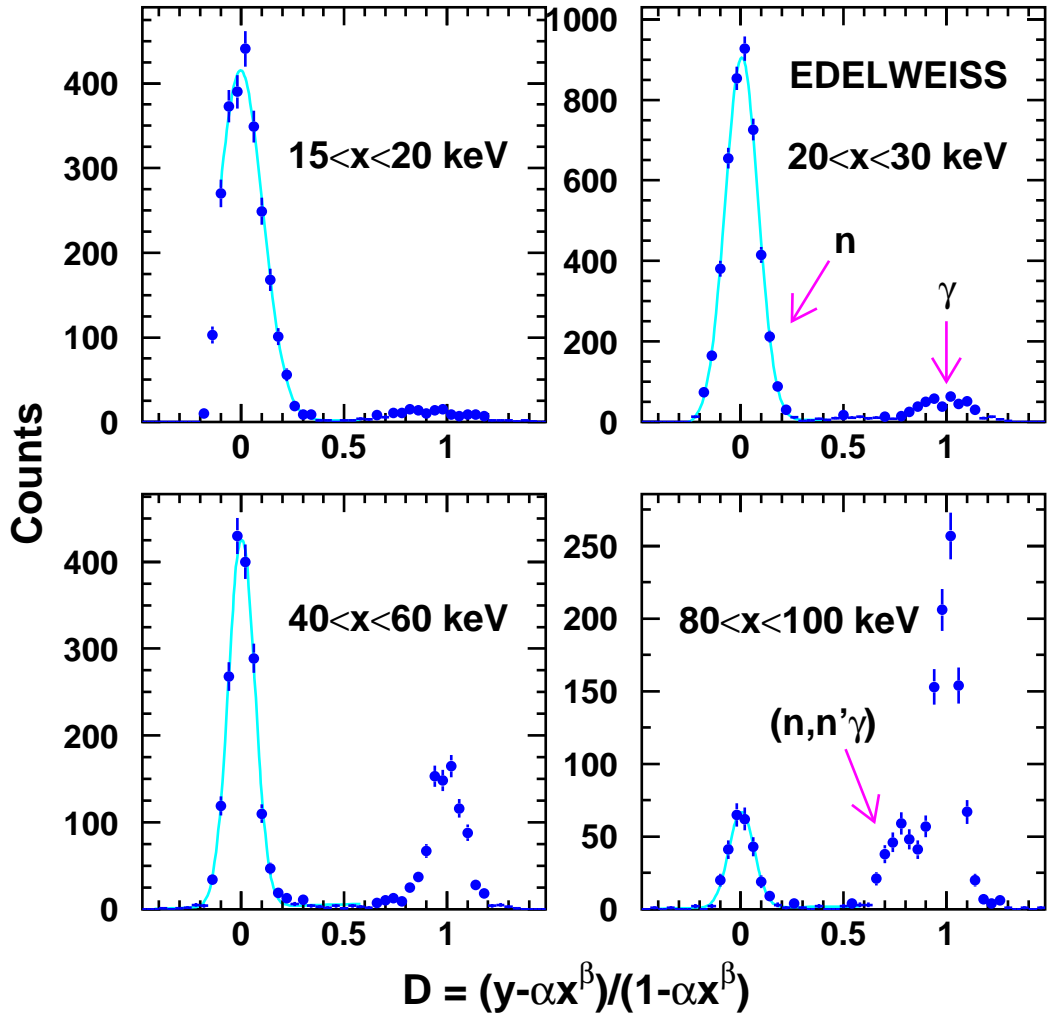}\end{center}
\caption{\label{cap:qprof}Experimental distributions of $D=(y-\alpha x^{\beta})/(1-\alpha x^{\beta})$
for different values of recoil energy interval $x$, for the sum of
the five EDELWEISS detectors used in the present work. Here, $\alpha=0.145$
and $\beta=0.201$.}
\end{figure}

\begin{figure}
\begin{center}\includegraphics[scale=0.8]{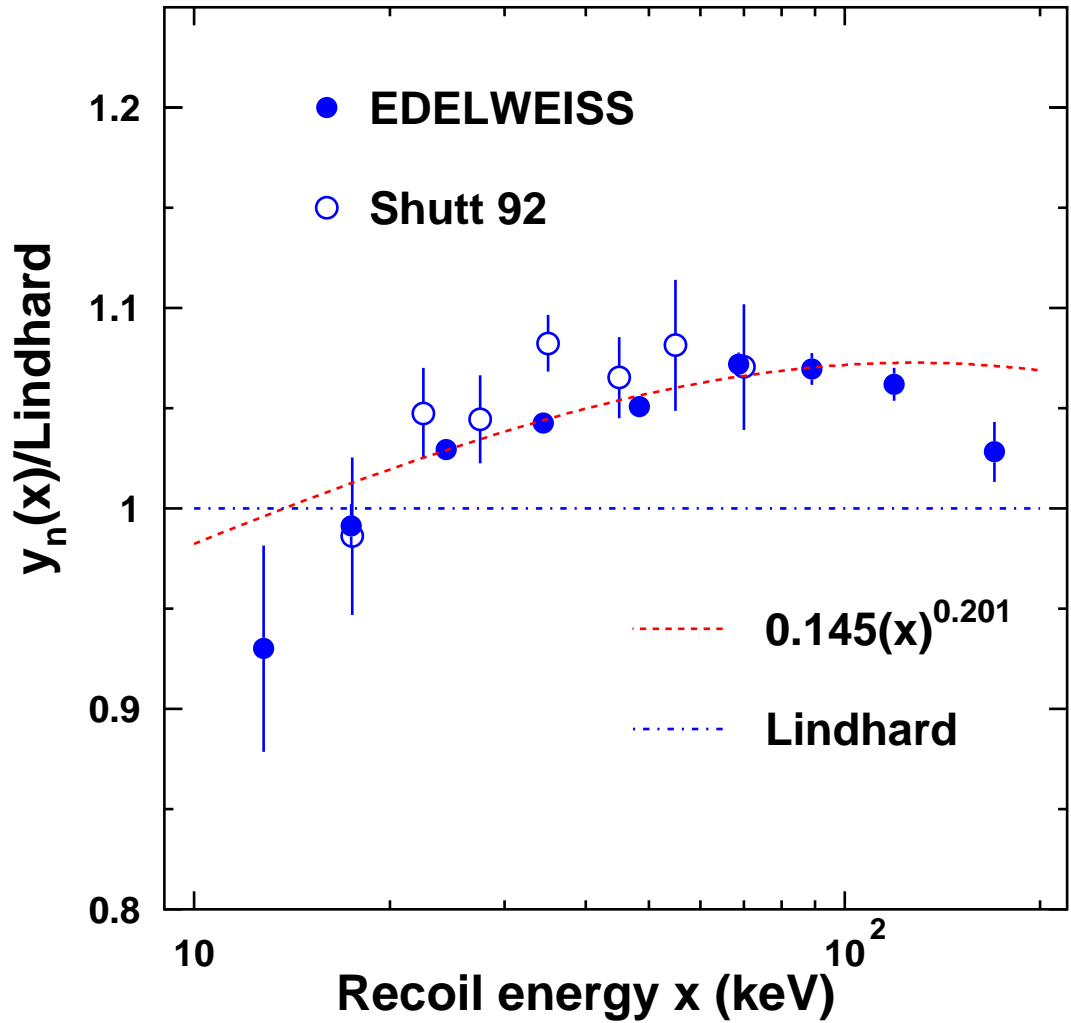}\end{center}
\caption{\label{cap:qedwshu}Ratio of the experimental values of $y_{n}$
divided by the Lindhard model for EDELWEISS (full circles) and the
data from Ref.~\cite{bib-shutt}. The same ratio is also shown for
the parameterization $y_{n}=0.145x^{0.201}$, corresponding to the 
best fit to the EDELWEISS
$y_{n}$ data (dashed line).}
\end{figure}

In Fig.~\ref{cap:yvsx}, there is a clear trend for $y$ to increase
with recoil energy for elastic nuclear scattering events. We define
$y_{n}(x)$ as the average $y$ values for this population, as a function
of $x$. The variations of $y_{n}(x)$ within the energy intervals
defined in Table~\ref{tab:q:} is not negligible at the level of
precision required here. Accordingly, this energy dependence is removed
by considering distributions of the variable $D$, defined as:\begin{equation}
D(x)=\frac{y(x)-\alpha x^{\beta}}{1-\alpha x^{\beta}}\label{eq:d}\end{equation}

where $\alpha$ and $\beta$ are chosen such as to describe best the
behavior of $y_{n}(x)$. Examples of $D(x)$ distributions recorded
in the neutron calibration runs for different intervals of recoil energy
$x$ are shown in fig.~\ref{cap:qprof}. In practice, the standard
values~\cite{edw2003,bib-edw2002,edw-calib} 
of $\alpha=0.16$ and $\beta=0.18$ are first used. The centroid
$D_{n}(x)$ of the $D$ distributions for elastic scattering events
is extracted for each bin of $x$. The $D_{n}(x)$ values are transformed
into values of $y_{n}(x)$ using the inverse of Eq.~\ref{eq:d}.
The results for the sum of the five EDELWEISS detectors under study
are listed in the third column of Table~\ref{tab-y}. They are also
shown in Fig.~\ref{cap:qedwshu}, where for clarity $y_{n}(x)$ is
divided by the corresponding prediction of the Lindhard theory, $Q_{ref}(x)$.
The $y_{n}(x)$ values are then least-square adjusted with the function
$\alpha'x^{\beta'}$, yielding values of $\alpha'=0.145$ and $\beta'=0.201$.
The entire procedure is then repeated, but this time using the new
values of $\alpha'$ and $\beta'$ in Eq.~\ref{eq:d}. With these
values, it is observed in Fig.~\ref{cap:qprof} that the new $D(x)$
distributions are well centered on $D_{n}=0$. However, the $y_{n}(x)$
values extracted from the new distributions are equal to the previous
ones, indicating that the procedure to remove the energy dependence
of $y_{n}(x)$ does not depend strongly on the precise choice of $\alpha$
and $\beta$.

In Fig.~\ref{cap:qedwshu}, the EDELWEISS data are compared to those
from Ref.~\cite{bib-shutt}. For both data sets, only statistical
errors are shown. Except for the values at $x$ = 35 keV, the experimental
results are compatible within the statistical uncertainties. However,
two important sources of bias have not been corrected yet, and will
be discussed in the following section. As Ref.~\cite{bib-shutt}
was the first report of the observation of quenching effects in heat-and-ionization
detector, it did not present the detailed information necessary for
evaluating these biases. In particular, there is no quantitative statement
concerning the precision of the $\gamma$-ray calibration, and no
discussion of the effect of multiple scattering. As it will be shown
in the following section, the multiple scattering of neutrons inside
the detectors decreases the apparent value of $y_{n}$. This effect
could explain why the data from the $\sim$180 g central volume of
the EDELWEISS detectors seem to lie systematically below those obtained
with the relatively more compact detector of Ref.~\cite{bib-shutt}.

\subsection{Multiple scattering and energy scale corrections}

\label{sub:corrections}In order to interpret the measured $y_{n}(x)$
values as $Q(E)/Q'(E)$, the data must be corrected for two systematic
effects: multiple scattering and the shift in energy scale when the
variable $x$ is transformed into a true recoil energy $E$. 

It was shown in Ref.~\cite{edw-calib} that the fact that there are
more than one neutron-nucleus interaction in a significant number
of neutron scattering events has the consequence of shifting down
the measured values of $y(x)$. For example, in an event where a single
neutron produces two nuclear recoils of energy $E_{1}$ and $E_{2}$,
the apparent quenching $Q_{eff}(E=E_{1}+E_{2})$ is equal to the energy-weighted
sum of $Q(E_{1})$ and $Q(E_{2})$. Depending on the relative values
of $E_{1}$ and $E_{2}$, the weighted sum will lie somewhere between
$Q(E/2)$ and $Q(E)$. As $dQ/dE$ $>$ 0, this means that the measured
value of $Q(E)$ decreases as the contribution from multiple scattering
increases.

This bias can be corrected with the help of Monte Carlo simulations.
In EDELWEISS, the results of simulations~\cite{bib-lem06} based on the computer codes
GEANT3 and MCNPX were compared and found to give consistent results
to within 1\%. The corresponding correction factors to be applied
to $y_{n}(x)$ are listed in Table~\ref{tab-y}. They are of the
order of 6 to 9\%. They are larger by 2 to 3\% compared with earlier
simulations performed for Ref.~\cite{edw-calib} with GEANT3. Since
then, problems with these simulations have been identified and solved,
leading to a better agreement with MCNPX. However, since this correction
relies entirely on simulations, the uncertainty on the correction
procedure is taken as a third of the correction.

The second correction corresponds to the application of Eq.~\ref{eq:ecor}.
It requires the knowledge of $Q'$, and thus must be evaluated iteratively,
first by evaluating it assuming $Q'$=1, and then replacing the derived
$Q'$ value in Eq.~\ref{eq:ecor}, and repeat until convergence.
Since $Q'$ is very close to unity, only one iteration was found necessary
to obtain stable corrections, listed in Table~\ref{tab-y}. The size
of the correction is approximately 2\%.

These two corrections are applied to the values of $y_{n}(x)$ in
the third column of Table~\ref{tab-y} in order to obtain the $Q(E)/Q'(E)$
values in the last column of the same table.

\subsection{Systematic uncertainties}

\label{sub:Syst}The seventh column of Table~\ref{tab-y} lists the
systematic uncertainties identified in addition to that associated
with the multiple scattering correction. The sources of systematic
biases considered here are those associated with the data selection
and with the uncertainties on the energy calibration of $E_{I}$ and
$E_{H}$, and on the value of $v$.

The energy calibration is reliable to within 1\%. In addition to the
test described in Sect.~\ref{sub:gcal}, it was checked that the
values of $y_{\gamma}$, the centroid of the $y(x)$ distributions
for electron recoils, stay within 1\% of unity from run to run, whether
with a neutron or $\gamma$-ray source or in the low-background runs.
For the contribution of $v$, the dependence of the heat gain as a
function of the applied voltage (Eq.~\ref{eq:ah}) was checked. Unaccounted
deviations of more than 0.1 V are excluded, resulting in an uncertainty
on $y_{n}$ of less than 1\%. The uncertainty on $y_{n}$ due to the
uncertainty on $\epsilon_{\gamma}$(Sect.~\ref{sub:Ion-meas}) is
less than 0.5\%.

Finally, the determination of the $y_{n}(x)$ values was repeated
for each of the five detectors individually. This not only checks
the effects of detector-to-detector fluctuations of the calibration,
but also the influence of the different experimental cuts. For example,
the lower cut on $E_{I}$ is 2.5 keV for GGA3, GSA1 and GSA3, and
is 3.5 keV for GeAl10 and GGA1. As discussed in Ref.~\cite{edw2003},
this corresponds to thresholds on recoil energy of approximately 11
and 14 keV, respectively. The effect of this cut can be observed on
Figs.~\ref{cap:yvsx} and~\ref{cap:qprof} as a decrease of efficiency
at the very lowest value of $y(x)$. For this reason, for $x$ values
below 20 keV, values of $D(x)$ below -0.1 are excluded from the analysis.
As a result, the fluctuations in $y_{n}(x)$ obtained from one
detector to another for $x<20$keV are larger than for most other
intervals (Table~\ref{tab:q:}). 

For each interval in $x$, the largest deviation of any detector from
the quoted $y_{n}(x)$ value is taken as a systematic uncertainty.
This value is added in quadrature to 1\% of the value of $y_{n}(x)$
to yield the systematic uncertainty listed in Table~\ref{tab-y}.
This conservative error tests possible deviations due to the technical
differences between the GeAl, GGA and GSA detectors. It also tests
the robustness of the measurement and in particular the reproducibility
of the calibration procedure.

This systematic error is added in quadrature with the uncertainty associated
with the multiple scattering correction and the statistical error,
resulting in the total error quoted in the last column of Table~\ref{tab-y}.
These three contributions represent typically relative errors of the
order of 3\%, 3\% and 0.5\% of the quoted $Q/Q'$ values.

Two further systematic checks were performed. In the first one, the
data selection and analysis procedure was applied to the simulated
data sample described in Sect.~\ref{sub:corrections}. The $y_{n}$
values extracted from the single scatter events are found to agree
with the $\alpha x^{\beta}$ parameterization that had been input
into the simulation, within a precision of 1\%. The systematic biases
due to the finite energy resolution and the selection procedure are
thus well within the quoted total systematic errors. In the second
test, values of $y_{n}$ were also extracted from inelastic scattering
events. In this test, the data were corrected for the presence of
an unquenched $\gamma$ ray with an energy $E_{\gamma}$ of either
13.3 or 68.8 keV. The remaining ionization and heat energies, $E_{H}-E_{\gamma}$
and $E_{I}-E_{\gamma}$, were input to Eqs.~\ref{eq:defx} and~\ref{eq:defy}.
Within statistical errors, the resulting $y(x)$ distribution shows
a band centered at the same values of $y_{n}(x)$ as the equivalent
distribution for elastic neutron scattering events. Because of the
reduced statistics ($\sim$1\% of the elastic scattering sample),
these data sets were not included in the final analysis.

The present method to evaluate $Q'$ assumes that 
the ionization quenching $Q$ does not change between 17 mK
(the temperatures at which $Q/Q'$ is measured)
and 77 K (the temperature for almost all of the $Q$ measurements). 
The only direct experimental test of this assumption is provided 
by the $Q$ measurements at 35 mK of Ref.~\cite{bib-sicane}. 
They are compatible with those performed at 77 K, within an experimental 
uncertainty of approximatively 5\%. However, there is no compelling
reason to suggest that $\epsilon_{\gamma}/\epsilon_{n}$ depends
on temperature, since the few percent variation of the gap in germanium
over this temperature range should affect equally both types of recoils. 
Therefore, the temperature dependence of $Q$ between 17 mK and 77 K
is assumed to be negligible.

\section{Q' results and discussion}

\begin{figure}
\begin{center}\includegraphics[scale=0.8]{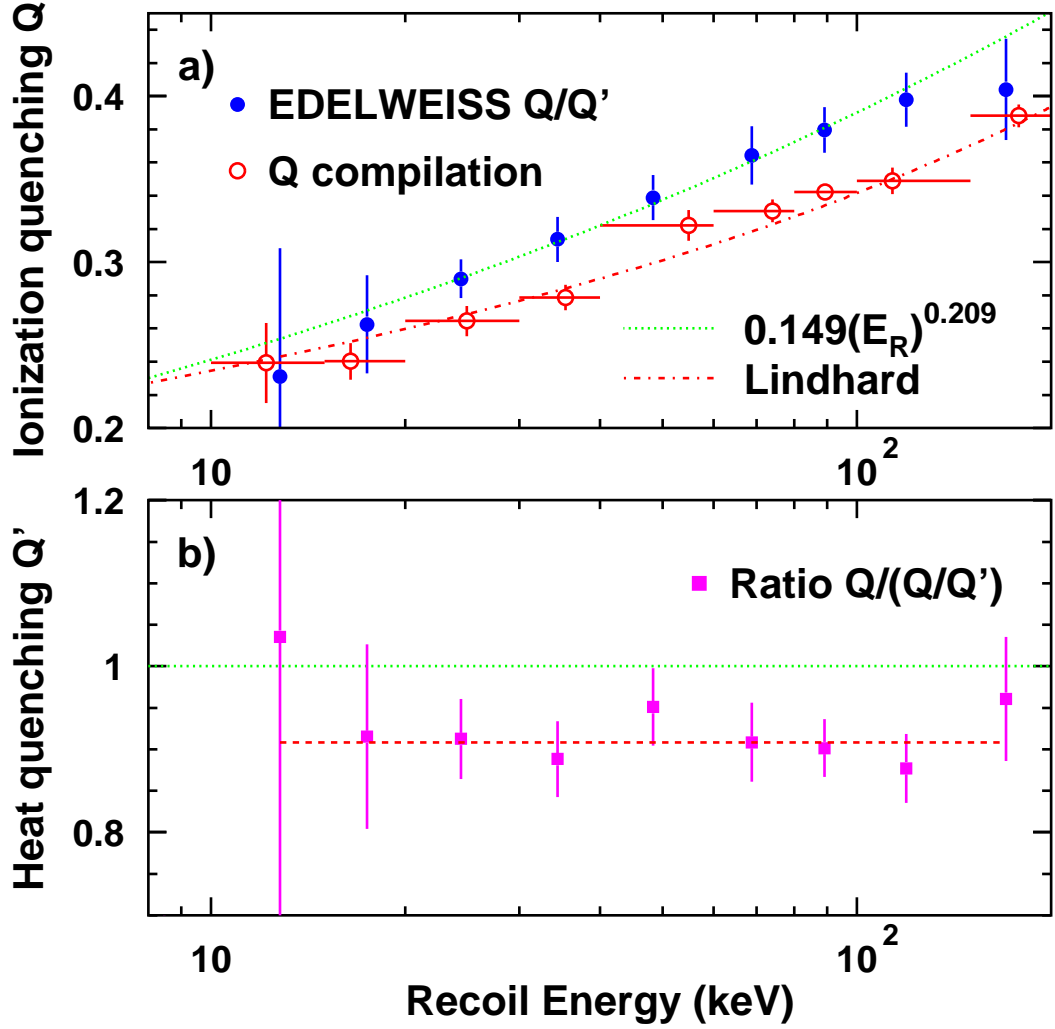}\end{center}
\caption{\label{cap:qprime}(a): values of $Q/Q'$ measured by EDELWEISS (full
circles) and values of $Q$ obtained in the compilation described
in Sect.~\ref{sec:Qrev} (open circles). The horizontal error bars
correspond to the range of the different energy intervals used in
the $Q$ and $Q/Q'$ analyses. The dot-dashed curve is the Lindhard
parameterization described in the text. The dotted curve represents
the best fit to the EDELWEISS $Q/Q'$ data ($Q/Q'=0.149(E_{R})^{0.209}$).
(b): $Q'$ values obtained from the ratio of these data. The dashed
line is the weighted average of all values ($Q'=0.91$).}
\end{figure}

Table~\ref{tab-qprime} summarizes the measurements of $Q(E)$ and
$Q(E)/Q'(E)$ from Sect.~\ref{sec:Qrev} and~\ref{sec:Q/Q'}, respectively.
These two data sets are also compared in Fig.~\ref{cap:qprime}a.
Because of the correction for multiple scattering and the relation
between $x$ and $E$, the best fit to the $Q/Q'$ data
($\alpha=0.149$ and $\beta=0.209$) differs slightly from the
$\alpha'$ and $\beta'$ values describing the $y_n$ data of 
Section~\ref{sub:ncal}.
This figure shows that the values of $Q(E)/Q'(E)$ are systematically larger than
those of $Q(E)$.
As a consequence, the resulting values of $Q'(E)$ deduced from their ratio
are systematically lower than unity 
(see Table ~\ref{tab-qprime} and Fig.~\ref{cap:qprime}b).
The weighted average of all measurements is 0.91.
As the uncertainty on the $Q'$ measurements at different energies
are correlated, it is not possible to reduce the error on this average.
However, it can be said that all values and errors between 20 and
100 keV are consistent with $Q'$ = 0.91 $\pm$ 0.05. Outside this
energy range, larger discrepancies are not excluded, as the measurements
are less precise. Typically, the contributions of the $Q(E)$ and
$Q(E)/Q'(E)$ measurements to the total uncertainty are $\pm$0.03
and $\pm$0.04, respectively. The latter error could be reduced by
improving our understanding of the influence of multiple scattering.

This determination of $Q'$ is the most precise measurement
of quenching in the bulk of a low-temperature bolometric detector
(Refs.~\cite{bib-zhou,bib-aless}, plotted on Fig.~\ref{cap:previous}).
It is three times more precise than the direct germanium measurement\cite{bib-sicane}.
It is the first precise measurement based on data other than Pb and
Po surface recoils from U/Th chain products.

The measurement indicates that the thermal responses of our detectors 
to nuclear and electron recoils are different. 
This is an incitation to evaluate with more precision the effects that could
be responsible. For example, simulations of the slowing down of 10
to 200 keV germanium recoils in germanium with the simplest version
of the program SRIM2003~\cite{bib-srim} provide some indications that 
a few percent
of the kinetic energy of the initial recoil can be trapped in defects
due to displacements of atoms in the matrix. However the actual fraction
depends considerably on the details of the simulation. How long these
defects persist in the out-of-equilibrium environment along the recoil
track is also a delicate question to address, especially in a low-temperature
semiconductor medium~\cite{bib-mdsim}.

It should be noted that the uncertainty related to the value of $Q'$
does not affect the evaluation of the efficiency for nuclear recoils,
nor the rejection capabilities of the EDELWEISS detectors. This is
because the bands for nuclear and electronic recoils are determined
for every detector using a neutron source calibration, with exactly
the same experimental conditions as in the physics runs. These bands
are defined in terms of $y(x)$ as observed in a given detector, and
do not rely on any prediction concerning $Q(E)$ and $Q'(E)$. 
However, the
value of $Q'$ does affect the interpretation of $x$ in terms of
absolute energy, and thus affects the overall efficiency of the detector
because of the presence of an experimental threshold. The effect of
a shift in energy scale on the limits on the WIMP-nucleon cross-section
set by the EDELWEISS data was studied in Ref.~\cite{edw2003}. In
this case, a 10\% shift of the energy scale was found to result in a 
10\% to 20\% shift on the cross-section limits, 
a small effect at the scale of the present sensitivity
of the experiments. In addition, the evaluation of both $Q$ and $Q'$
presented here are among the most precise quenching measurements
for any detector used in the direct search for dark matter. In contrast,
little is available concerning the scintillation quenching of iodine
in NaI~\cite{bib-qiode}. The experimental data on the quenching in Xe are only now just
starting to converge and to cover the low energy range
relevant for dark matter searches~\cite{bib-qxe}. 
The wealth of available data on germanium and the consistency of the 
results make it one of the most reliable technique for detecting and 
identifying nuclear recoils.

\section{Conclusion}

The heat quenching factor (the ratio of the heat signals produced
by nuclear and electron recoils of equal energy) of the heat-and-ionization
germanium bolometers used by the EDELWEISS collaboration has been
measured. 
It is shown that the calibration of these detectors with neutrons provides a
measurement of the ratio $Q/Q'$, where $Q$ and $Q'$ are the quenching factors 
of the ionization yield and heat measurement, respectively. 
Consequently, the existing direct measurements of $Q$ have been reviewed 
and the resulting values have been divided by the EDELWEISS $Q/Q'$ measurements.
The resulting heat quenching factor for germanium recoil energies
between 20 and 100 keV is $Q'$ = 0.91 $\pm$ 0.03 $\pm$ 0.04, where
the two errors are the contributions from the $Q$ and $Q/Q'$ measurements,
respectively. The evaluation of both $Q$ and $Q'$ presented here
are among the most precise quenching measurements for any detector
used in the direct search for dark matter.

\begin{acknowledgments}
The help of the technical staff of the Laboratoire 
Souterrain de Modane and the participant laboratories is 
gratefully acknowledged. This work has been partially 
supported by the EEC Applied Cryodetector network 
(Contract No. HPRN-CT-2002-00322) and the ILIAS integrating activity 
(Contract No. RII3-CT-2003-506222). 
\end{acknowledgments}

\newpage

\begin{table*}
\begin{center}\begin{tabular}{|c|c|c|c|c|c|c|}
\hline 
$E_{R}$&
Refs.&
$Q/Q_{ref}$&
$\sqrt{\frac{\chi^{2}}{(N-1)}}$&
$<E_{R}>$&
$Q_{ref}(<E_{R}>)$&
$Q(<E_{R}>)$\tabularnewline
(keV)&
&
&
&
(keV)&
&
\tabularnewline
\hline
\hline 
5-10&
\cite{bib-messous}&
1.05 $\pm$ 0.08&
-&
8.6&
0.230&
0.241 $\pm$ 0.019\tabularnewline
\hline 
10-15&
\cite{bib-chasman2,bib-messous}&
0.99 $\pm$ 0.10&
2.01&
12.2&
0.241&
0.239 $\pm$ 0.024\tabularnewline
\hline 
15-20&
\cite{bib-chasman2,bib-messous}&
0.95 $\pm$ 0.04&
0.34&
16.4&
0.252&
0.240 $\pm$ 0.011\tabularnewline
\hline 
20-30&
\cite{bib-chasman1,bib-chasman2,bib-messous}&
0.98 $\pm$ 0.03&
1.29&
24.9&
0.269&
0.265 $\pm$ 0.009\tabularnewline
\hline 
30-40&
\cite{bib-chasman1,bib-messous}&
0.98 $\pm$ 0.03&
1.34&
35.4&
0.284&
0.279 $\pm$ 0.008\tabularnewline
\hline 
40-60&
\cite{bib-chasman1,bib-chasman2,bib-baudis}&
1.05 $\pm$ 0.03&
0.48&
54.9&
0.306&
0.322 $\pm$ 0.009\tabularnewline
\hline 
60-80&
\cite{bib-chasman1,bib-baudis,bib-sicane}&
1.03 $\pm$ 0.02&
0.71&
74.0&
0.323&
0.331 $\pm$ 0.007\tabularnewline
\hline 
80-100&
\cite{bib-chasman1,bib-baudis,bib-sicane}&
1.02 $\pm$ 0.01&
0.31&
89.3&
0.334&
0.342 $\pm$ 0.005\tabularnewline
\hline 
100-150&
\cite{bib-chasman1,bib-baudis,bib-sicane,bib-chasman2}&
1.00 $\pm$ 0.02&
2.21&
113.6&
0.350&
0.349 $\pm$ 0.008\tabularnewline
\hline 
150-200&
\cite{bib-baudis}&
1.01 $\pm$ 0.02&
-&
178.2&
0.384&
0.388 $\pm$ 0.007\tabularnewline
\hline
\end{tabular}\end{center}

\caption{\label{tab:q:}Average of measurements of $Q$, the ionization quenching
of Ge recoils in Ge, as a function of the recoil energy $E_{R}$.}
\end{table*}

\begin{table*}[ht]
\begin{center}\begin{tabular}{|c|c|c|c|c|c|c|c|}
\hline 
$E_{R}$&
$<E_{R}>$&
$y_{n}(x)$&
Stat.&
Multiple scattering&
$x\rightarrow E$&
Calib.&
$Q/Q'$\tabularnewline
(keV)&
(keV)&
&
error&
correction&
(Eq. 12)&
error&
\tabularnewline
\hline
\hline 
10-15&
12.8&
0.226&
0.014&
1.039 $\pm$ 0.013&
0.985&
0.017&
0.231 $\pm$ 0.077\tabularnewline
\hline 
15-20&
17.4&
0.252&
0.003&
1.059 $\pm$ 0.020&
0.983&
0.008&
0.263 $\pm$ 0.029\tabularnewline
\hline 
20-30&
24.4&
0.276&
0.001&
1.070 $\pm$ 0.023&
0.983&
0.003&
0.290 $\pm$ 0.012\tabularnewline
\hline 
30-40&
34.4&
0.295&
0.001&
1.083 $\pm$ 0.028&
0.982&
0.004&
0.314 $\pm$ 0.014\tabularnewline
\hline 
40-60&
48.4&
0.315&
0.001&
1.096 $\pm$ 0.032&
0.982&
0.003&
0.339 $\pm$ 0.013\tabularnewline
\hline 
60-80&
68.7&
0.341&
0.002&
1.088 $\pm$ 0.029&
0.981&
0.005&
0.364 $\pm$ 0.017\tabularnewline
\hline 
80-100&
89.1&
0.357&
0.003&
1.084 $\pm$ 0.028&
0.980&
0.004&
0.380 $\pm$ 0.014\tabularnewline
\hline 
100-150&
119.3&
0.375&
0.003&
1.083 $\pm$ 0.028&
0.979&
0.006&
0.398 $\pm$ 0.016\tabularnewline
\hline 
150-200&
170.1&
0.391&
0.006&
1.057 $\pm$ 0.019&
0.978&
0.012&
0.404 $\pm$ 0.031\tabularnewline
\hline
\end{tabular}\end{center}

\caption{\label{tab-y}Measurements of $Q/Q'$ for germanium recoils in germanium
obtained from the neutron calibration of the EDELWEISS detectors GeAl10,
GGA1, GGA3, GSA1 and GSA3.}
\end{table*}

\begin{table}[ht]
\begin{center}\begin{tabular}{|c|c|c|c|c|}
\hline 
$E_{R}$&
$<E_{R}>$&
$Q(E)/Q'(E)$&
$Q$&
$Q'$\tabularnewline
(keV)&
(keV)&
&
&
\tabularnewline
\hline
\hline 
10-15&
12.8&
0.231 $\pm$ 0.077&
0.239 $\pm$ 0.024&
1.04 $\pm$ 0.36\tabularnewline
\hline 
15-20&
17.4&
0.263 $\pm$ 0.029&
0.240 $\pm$ 0.011&
0.91 $\pm$ 0.11\tabularnewline
\hline 
20-30&
24.4&
0.290 $\pm$ 0.012&
0.265 $\pm$ 0.009&
0.91 $\pm$ 0.05\tabularnewline
\hline 
30-40&
34,4&
0.314 $\pm$ 0.014&
0.279 $\pm$ 0.008&
0.89 $\pm$ 0.05\tabularnewline
\hline 
40-60&
48.4&
0.339 $\pm$ 0.013&
0.322 $\pm$ 0.009&
0.95 $\pm$ 0.05\tabularnewline
\hline 
60-80&
68.7&
0.364 $\pm$ 0.017&
0.331 $\pm$ 0.007&
0.91 $\pm$ 0.05\tabularnewline
\hline 
80-100&
89.1&
0.380 $\pm$ 0.014&
0.342 $\pm$ 0.005&
0.90 $\pm$ 0.03\tabularnewline
\hline 
100-150&
119.3&
0.398 $\pm$ 0.016&
0.349 $\pm$ 0.008&
0.88 $\pm$ 0.04\tabularnewline
\hline 
150-200&
170.1&
0.404 $\pm$ 0.031&
0.388 $\pm$ 0.007&
0.96 $\pm$ 0.07\tabularnewline
\hline
\end{tabular}\end{center}

\caption{\label{tab-qprime}Summary of the results of the compilation of $Q$
measurements, of the EDELWEISS $Q/Q'$ measurement, and the deduced
values of $Q'$ as a function of recoil energy.}
\end{table}

\clearpage
\newpage

\end{document}